\documentclass[12pt]{article}

\def\case#1#2{{\textstyle{#1\over #2}}}
\def\ap{a^{\dagger}}
\def\bbeta{\bar{\beta}}

\title{
\hfill{\normalsize ULB/229/CQ/01/7}\\
\vspace{1cm}
\boldmath Generalized Coherent States Associated with the
$C_{\lambda}$-Extended Oscillator}

\author{C. Quesne\\
{\small Physique Nucl\'eaire Th\'eorique et Physique Math\'ematique, Universit\'e
Libre de Bruxelles,}\\
{\small Campus de la Plaine CP229, Boulevard du Triomphe, B-1050
Brussels, Belgium}\\   
{\small E-mail: cquesne@ulb.ac.be }}

\date{}

\begin{document}
\baselineskip=20pt plus 1pt minus 1pt
\maketitle

\begin{abstract}
Two new types of coherent states associated with the 
$C_{\lambda}$-extended oscillator, where $C_{\lambda}$ is the cyclic group of order
$\lambda$, are introduced. They satisfy a unity resolution relation in the
$C_{\lambda}$-extended oscillator Fock space (or in some subspace thereof) and give
rise to Bargmann representations of the latter, wherein the generators of the
$C_{\lambda}$-extended oscillator algebra are realized as differential-operator-valued
matrices.
\end{abstract}

\section{Introduction} 

Coherent states~(CS) of the harmonic oscillator are known to have properties similar to
those of the classical radiation field. In contrast, generalized CS associated with various 
algebras may have some nonclassical properties.\par
%
%
In the present communication, we shall consider some generalized CS,~\cite{cq00,cq01}
which may be associated with the recently introduced $C_{\lambda}$-extended
oscillator.~\cite{cq98} The latter may be considered as a deformed oscillator with  a
Z$_{\lambda}$-graded Fock space and has proved very useful in the context of
supersymmetric quantum mechanics and some of its variants.~\cite{cq98,cq99}\par
%
%
The main mathematical feature of these new generalized CS is that they satisfy Klauder's
minimal set of conditions for generalized CS,~\cite{klauder} including the existence of a
unity resolution relation, property that is not shared by generic generalized CS.\par
%
%
\section{\boldmath The $C_{\lambda}$-Extended Oscillator}

The $C_{\lambda}$-extended oscillator Hamiltonian is defined by~\cite{cq98}
\begin{equation}
  H_0 = \case{1}{2} \left\{a, \ap\right\},  \label{eq:hamiltonian}  
\end{equation}
where the creation and annihilation operators $\ap$, $a$ satisfy the commutation
relation
\begin{equation}
  \left[a, \ap\right] = I + \sum_{\mu=1}^{\lambda-1} \kappa_{\mu} T^{\mu} = I +
  \sum_{\mu=0}^{\lambda-1} \alpha_{\mu} P_{\mu}.  \label{eq:alg-commut}  
\end{equation}
Here $T = (T^{\dagger})^{-1}$ denotes the generator of the cyclic group 
$C_{\lambda} = \{T, T^2, \ldots, T^{\lambda-1}, T^{\lambda} = I\}$ (or, more
precisely, the generator of a unitary representation thereof), and $P_{\mu} =
\frac{1}{\lambda} 
\sum_{\nu=0}^{\lambda-1} \exp(-2\pi i\mu\nu/\lambda)\, T^{\nu}$, $\mu = 0$, 1,
\ldots,~$\lambda-1$, are the projectors on the $\lambda$ inequivalent unitary
irreducible matrix representations of $C_{\lambda}$. The parameters $\kappa_{\mu}$
are complex and satisfy the conditions $\kappa_{\mu}^* = \kappa_{\lambda - \mu}$,
whereas the $\alpha_{\mu}$'s are real and such that $\sum_{\mu=0}^{\lambda-1}
\alpha_{\mu} =0$. The commutation rule for $T$ (or $P_{\mu}$) and $\ap$ is given by
\begin{equation}
  \ap T = e^{-2\pi {\rm i}/\lambda} T \ap \qquad {\rm or} \qquad \ap P_{\mu} =
  P_{\mu+1} \ap.
\end{equation}
The $C_{\lambda}$-extended oscillator algebra also contains a number operator $N$
such that 
\begin{equation}
  \left[N, \ap\right] = \ap, \qquad [N, T] = 0 \qquad {\rm or} \qquad [N, P_{\mu}] = 0. 
\end{equation}
\par
%
%
When $T$ (or $P_{\mu}$) is realized in terms of $N$, i.e., $T = \exp(2\pi {\rm i} N/
\lambda)$, the $C_{\lambda}$-extended oscillator algebra becomes a generalized
deformed oscillator algebra~(GDOA). In the Fock representation of the latter, there exists
a vacuum state $|0\rangle$, such that $a |0\rangle = N |0\rangle = 0$, and the
remaining basis states $|n\rangle$, $n=1$, 2,~\ldots, can be obtained from it by
successive applications of $\ap$. The Fock space $\cal F$ is ${\rm
Z}_{\lambda}$-graded: ${\cal F} = \sum_{\mu=0}^{\lambda-1} {\cal F}_{\mu}$, where
${\cal F}_{\mu} = \{\, |n\rangle = |k\lambda + \mu\rangle \mid k=0, 1, 2, \ldots\, \}$,
and $P_{\mu}$ projects on ${\cal F}_{\mu}$.\par
%
%
The $C_{\lambda}$-extended oscillator Hamiltonian $H_0$, defined in
Eq.~(\ref{eq:hamiltonian}), has a spectrum made of $\lambda$ infinite sets of equally
spaced levels, belonging to ${\cal F}_{\mu}$, $\mu=0$, 1, \ldots,~$\lambda-1$,
respectively. Its spectrum generating algebra~(SGA) is generated by the operators $J_+
= (1/\lambda) (\ap)^{\lambda}$, $J_- = (1/\lambda) a^{\lambda}$, $J_0 =
(1/\lambda) H_0$, and is a $C_{\lambda}$-extended polynomial deformation of su(1,1)
for $\lambda > 2$, while it reduces to su(1,1) for  $\lambda=2$.~\cite{cq00}\par
%
%
{}For  $\lambda = 2$, the GDOA corresponding to the $C_{\lambda}$-extended oscillator
algebra reduces to the Calogero-Vasiliev algebra, characterized by the commutation
relation $[a, \ap] = I + \alpha_0 K$, where $K = T = (-1)^N$.~\cite{vasiliev} For
$\alpha_0 = p-1$, the latter becomes the paraboson algebra of order
$p$.~\cite{chaturvedi}\par
%
%
\section{\boldmath Family of Coherent States $|z; \mu; \alpha\rangle$}

The coherent states $|z; \mu; \alpha\rangle$, where $z \in {\rm C}$, $\alpha \in \{0,
1, \ldots, [\lambda/2]\}$, $\mu \in \{0, 1, \ldots, \lambda - \alpha - 1\}$, are the
solutions of the equation~\cite{cq01}
\begin{equation}
  \left[a^{\lambda - \alpha} - z \left(\ap\right)^{\alpha}\right] |z; \mu; \alpha\rangle
  = 0.  \label{eq:CS1-def}
\end{equation}
As special cases, they include (i) the annihilation-operator CS of the
$C_{\lambda}$-extended oscillator SGA, corresponding to $\alpha=0$ (i.e., $J_- |z; \mu;
0\rangle = (z/\lambda) |z; \mu; 0\rangle$),~\cite{cq00} and (ii) the
displacement-operator or Perelomov su(1,1) CS, corresponding to $\lambda=2$,
$\mu=0$, $\alpha=1$ (i.e., $(a - z \ap) |z; 0; 1\rangle =0$, where $|z; 0; 1\rangle
\propto \exp(z J_+) |0\rangle$).~\cite{perelomov}\par
%
%
The states $|z; \mu; \alpha\rangle$ belong to the subspace ${\cal F}_{\mu}$ of
$\cal F$ and can therefore be expanded in the number states $|k \lambda +
\mu\rangle$, $k=0$, 1,~\ldots, as
\begin{equation}
  |z; \mu; \alpha\rangle = \left[N^{(\alpha)}_{\mu}(|z|)\right]^{-1/2}
  \sum_{k=0}^{\infty} c'_k(z, z^*; \mu; \alpha) z^k |k\lambda + \mu\rangle,  
\end{equation}
where the coefficients $c'_k(z, z^*; \mu; \alpha)$ are given in Ref.~[2] and the
normalization coefficient can be expressed as
\begin{eqnarray}
  \lefteqn{N^{(\alpha)}_{\mu}(|z|) =} \nonumber \\[4pt]
  && {}_{\alpha}F_{\lambda - \alpha - 1} \left(\bbeta_{\mu+1},
  \ldots, \bbeta_{\mu+\alpha}; \bbeta_1 + 1, \ldots, \bbeta_{\mu} + 1, 
  \bbeta_{\mu+\alpha+1}, \ldots, \bbeta_{\lambda-1}; y\right). \label{eq:N}
\end{eqnarray}
In Eq.~(\ref{eq:N}), the variable $y$ is defined by $y \equiv |z|^2/\lambda^{\lambda -
2\alpha}$, while the constants $\bbeta_{\mu}$ are given in terms of the parameters
$\alpha_{\nu}$ of the $C_{\lambda}$-extended oscillator algebra  through the relations
$\bbeta_{\mu} \equiv (\beta_{\mu} + \mu)/\lambda$, $\beta_{\mu} \equiv
\sum_{\nu=0}^{\mu-1} \alpha_{\nu}$. From this, we conclude that the states $|z; \mu;
\alpha\rangle$ are normalizable on the complex plane for any $\alpha \in \{0, 1, \ldots,
[(\lambda - 1)/2]\}$ and on the unit disc for $\lambda$ even, $ \alpha =
\lambda/2$.\par
%
%
In addition, the CS $|z; \mu; \alpha\rangle$ are continous in the label $z$, i.e., $|z
-z'| \to 0 \Rightarrow \bigl||z; \mu; \alpha\rangle - |z'; \mu; \alpha\rangle\bigr|^2 \to
0$, and for appropriate values of the algebra parameters, they satisfy a
unity resolution relation with a positive measure, 
\begin{equation}
  \int d^2z\,  h^{(\alpha)}_{\mu} (y) |z; \mu; \alpha) (z; \mu; \alpha| = I_{\mu}, 
  \label{eq:resol-mu} 
\end{equation}
in ${\cal F}_{\mu}$ (with $I_{\mu}$ the unit operator in ${\cal F}_{\mu}$).
Alternatively, we may write
\begin{equation}
  \sum_{\mu=0}^{\lambda-1} \int d^2z\,  h^{(\alpha)}_{\mu} (y) |z; \mu; \alpha) (z;
  \mu; \alpha| = I  \label{eq:resol}
\end{equation}
in $\cal F$. In Eqs.~(\ref{eq:resol-mu}) and~(\ref{eq:resol}), $|z; \mu; \alpha) \equiv
\bigl[N^{(\alpha)}_{\mu}(|z|)\bigr]^{1/2} |z; \mu; \alpha\rangle$ denotes an
unnormalized CS and the general form of $h^{(\alpha)}_{\mu} (y)$ is conjectured to be
given in terms of a Meijer $G$-function as
\begin{equation}
  h^{(\alpha)}_{\mu}(y) = A^{(\alpha)}_{\mu} G^{\lambda - \alpha, 0}_{\alpha, \lambda
  - \alpha} \left(y\, \Bigg| 
      \begin{array}{c}
         \bbeta_{\mu+1} - 1, \ldots, \bbeta_{\mu+\alpha} - 1\\[0.1cm]
         0, \bbeta_1, \ldots, \bbeta_{\mu}, \bbeta_{\mu+\alpha+1} - 1, \ldots, 
              \bbeta_{\lambda-1} - 1
      \end{array}\right),  \label{eq:halphamu}
\end{equation}
where $A^{(\alpha)}_{\mu}$ is some numerical coefficient. Eq.~(\ref{eq:halphamu}) has
actually been proved for $\alpha \ne \lambda/2$ and any $\lambda$, and for $\alpha =
\lambda/2$ and $\lambda = 2$, 4, 6.~\cite{cq01}\par
%
%
Since the CS $|z; \mu; \alpha)$ form a complete (in fact, an overcomplete) set in
${\cal F}_{\mu}$ for some appropriate choice of the algebra parameters if $z$ runs over
the complex plane for $\alpha \le [(\lambda-1)/2]$ or over the unit disc for $\alpha =
\lambda/2$ and $\lambda$ even, we may associate with such a set a realization of ${\cal
F}_{\mu}$ as a Hilbert space ${\cal B}^{(\alpha)}_{\mu}$ of entire analytic functions .
This gives rise to so-called Bargmann representations~\cite{bargmann}  of the
$C_{\lambda}$-extended oscillator, wherein the operators $X$ defined in ${\cal
F}_{\mu}$ become differential operators ${\cal X}^{(\alpha)}_{\mu}$. For the SGA
generators $J_+$, $J_-$, and
$J_0$, for instance, we get
\begin{eqnarray}
  {\cal J}^{(\alpha)}_{+\mu} & = & \lambda^{\alpha-1} z 
         \prod_{\nu=\mu+1}^{\mu+\alpha} \left(z \frac{\partial}{\partial z} +
         \bbeta_{\nu}\right),  \label{eq:B1-J+} \\
  {\cal J}^{(\alpha)}_{-\mu} & = & \lambda^{\lambda-\alpha-1} \left[ 
         \prod_{\nu=1}^{\mu} \left(z \frac{\partial}{\partial z} + \bbeta_{\nu} + 1\right) 
         \right] \left[\prod_{\nu=\mu+\alpha+1}^{\lambda-1} \left(z
         \frac{\partial}{\partial z} + \bbeta_{\nu}\right)\right] \frac{\partial}{\partial z}, 
         \label{eq:B1-J-} \\
  {\cal J}^{(\alpha)}_{0\mu} & = & z \frac{\partial}{\partial z} + \frac{1}{2} 
         (\bbeta_{\mu} + \bbeta_{\mu+1}).  \label{eq:B1-J0}
\end{eqnarray}
\par
%
%
When dealing with the whole Fock space $\cal F$, we may consider as complete set the
collection of CS $|z; \mu; 0)$, $\mu=0$, 1, \ldots,~$\lambda-1$, and realize
$\cal F$ as a Hilbert space ${\cal B}^{(0)} = \sum_{\mu=0}^{\lambda-1} \oplus {\cal
B}^{(0)}_{\mu}$ of entire analytic functions. It is then convenient to introduce vector
CS~\cite{deenen} defined as row vectors, $||z; 0)) \equiv \bigl(|z; 0; 0), |z; 1; 0), \ldots,
|z; \lambda-1; 0)\bigr)$, the corresponding bras $((z; 0||$ being column vectors. In such
a case, the operators defined in $\cal F$ are represented by some operator-valued
$\lambda \times \lambda$ matrices.\par
%
%
\section{\boldmath Eigenstates $|z\rangle$ of the Annihilation Operator $a$}

The eigenstates of the $C_{\lambda}$-extended oscillator annihilation operator
$a$, defined by~\cite{cq01}
\begin{equation}
  a |z\rangle = z |z\rangle,  \label{eq:CS2-def}
\end{equation}
where $z \in {\rm C}$, include as special case the paraboson CS,~\cite{sharma}
corresponding to $\lambda=2$.\par
%
%
They can be constructed as linear combinations of the CS $|\omega; \mu;
0\rangle$, $\mu=0$, 1, \ldots,~$\lambda-1$, considered in Sec.~3, where one
replaces $z$ by $\omega \equiv z^{\lambda}$,
\begin{equation}
  |z\rangle = [{\cal N}(|z|)]^{-1/2}\sum_{\mu=0}^{\lambda-1} d'_{\mu}\left(z,
  z^*\right) \left(\frac{z}{\sqrt{\lambda}}\right)^{\mu} |\omega; \mu; 0\rangle,
  \label{eq:CS2-ansatz} 
\end{equation}
with $d'_{\mu}(z, z^*) = \bigl[N^{(0)}_{\mu}(|\omega|)/\prod_{\nu=1}^{\mu}
\bbeta_{\nu}\bigr]^{1/2}$. The normalization coefficient ${\cal N}(|z|)$ is given
by
\begin{equation}
  {\cal N}(|z|) =\sum_{\mu=0}^{\lambda-1} {}_0F_{\lambda-1} \left(\bbeta_1 + 1,
  \ldots, \bbeta_{\mu} + 1, \bbeta_{\mu+1}, \ldots, \bbeta_{\lambda-1}; t^{\lambda}
  \right) \frac{t^{\mu}}{\prod_{\nu=1}^{\mu} \bbeta_{\nu}}, 
\end{equation}
where $t \equiv |z|^2/\lambda$. It is therefore straightforward to see that the CS
$|z\rangle$ are normalizable on the complex plane and continuous in their label
$z$.\par
%
%
They satisfy a unity resolution relation of unusual form,~\cite{cq01}
\begin{equation}
  \sum_{\mu=0}^{\lambda-1} \int d^2z\, h_{\mu}(t) |z_{\mu}) (z_{\mu}| =
  \sum_{\mu=0}^{\lambda-1} \int d^2z\, g_{\mu}(t) |z)
  \bigl( z e^{2\pi {\rm i} \mu/\lambda} \bigr| = I,  \label{eq:CS2-resol2}
\end{equation}
generalizing the known one for paraboson CS.~\cite{sharma} Here $|z) = [{\cal
N}(|z|)]^{1/2} |z\rangle$ denotes an unnormalized CS, $|z_{\mu})$ its component in
${\cal F}_{\mu}$, and
\begin{equation}
  h_{\mu}(t) = \lambda^{\lambda} \left(\prod_{\nu=1}^{\mu} \bbeta_{\nu}\right)
  t^{\lambda - \mu - 1} h^{(0)}_{\mu}(t^{\lambda}), \qquad
  g_{\mu}(t) = \frac{1}{\lambda} \sum_{\nu=0}^{\lambda-1} e^{2\pi {\rm i}
         \mu \nu/\lambda} h_{\nu}(t), 
\end{equation}
where $h^{(0)}_{\mu}(t^{\lambda})$ is defined in Eq.~(\ref{eq:halphamu}). It can
actually be shown that this nondiagonal CS resolution of unity is entirely equivalent to the
diagonal one valid for the set of CS $|z; \mu; 0)$, $\mu=0$, 1, \ldots,~$\lambda-1$,
and given in Eq.~(\ref{eq:resol}). 
\par
%
%
Instead of the set of CS $|z; \mu; 0)$, $\mu=0$, 1, \ldots,~$\lambda-1$, we may
use $|z_{\mu})$, $\mu=0$, 1, \ldots,~$\lambda-1$, as complet set in $\cal F$
and introduce a vector notation again, $||z)) \equiv \bigl(|z_0), |z_1), \ldots,
|z_{\lambda-1})\bigr)$. In the corresponding Bargmann space $\cal B$, the
operators $N$, $\ap$, $a$ of the $C_{\lambda}$-extended oscillator algebra take a very
simple form, namely
\begin{equation}
  {\cal N} = z \frac{\partial}{\partial z} {\cal I},
\end{equation}
\begin{equation}
  {\cal A}^{\dagger} = \left(\begin{array}{ccccc}
          0 & 0 & \ldots & 0 & z \\
          z & 0 & \ldots & 0 & 0 \\
          0 & z & \ldots & 0 & 0 \\
          \vdots & \vdots & \ddots & \vdots & \vdots \\
          0 & 0 & \ldots & z & 0
        \end{array}\right), \qquad
  {\cal A} = \left(\begin{array}{ccccc}
          0 & \frac{\partial}{\partial z} + \frac{\beta_1}{z} & 0 & \ldots & 0 \\
          0 & 0 & \frac{\partial}{\partial z} + \frac{\beta_2}{z} & \ldots & 0 \\
          \vdots & \vdots & \vdots & \ddots & \vdots \\
          0 & 0 & 0 & \ldots & \frac{\partial}{\partial z} + \frac{\beta_{\lambda-1}}{z}
                \\
          \frac{\partial}{\partial z} & 0 & 0 & \ldots & 0
        \end{array}\right), 
\end{equation} 
where $\cal I$ denotes the $\lambda \times \lambda$ unit matrix.\par
%
%
\section{Conclusion}

In the present contribution, we presented the mathematical properties of some new CS
associated with the $C_{\lambda}$-extended oscillator.\par
%
%
Such states have also some interesting physical properties. In Refs.~[1, 2],
we indeed investigated some of their characteristics relevant to quantum optics, such as
their statistical and squeezing properties, for a wide range of parameters and from both
viewpoints of real and dressed photons. Their nonclassical features for some parameter
ranges were clearly demonstrated.\par
%
%
\section*{Acknowledgments}

The author is a Research Director of the National Fund for Scientific Research
(FNRS), Belgium.          
%
%


\begin{thebibliography}{99}

\bibitem{cq00}C. Quesne, {\em Phys. Lett.} A {\bf 272}, 313 (2000); {\em ibid.}
{\bf 275}, 313 (2000).

\bibitem{cq01}C. Quesne, {\em Ann. Phys. (N.Y.)}, in press.

\bibitem{cq98}C. Quesne and N. Vansteenkiste, {\em Phys. Lett.} A {\bf 240}, 21
(1998).

\bibitem{cq99}C. Quesne and N. Vansteenkiste, {\em Helv. Phys. Acta} {\bf 72}, 71
(1999); {\em Int. J. Theor. Phys.} {\bf 39}, 1175 (2000).

\bibitem{klauder}J.R. Klauder, {\em J. Math. Phys.} {\bf 4}, 1055 (1963); {\em ibid.}
{\bf 4}, 1058 (1963).

\bibitem{vasiliev}M.A. Vasiliev, {\em Int. J. Mod. Phys.} A {\bf 6}, 1115 (1991).

\bibitem{chaturvedi}S. Chaturvedi and V. Srinivasan, {\em Phys. Rev.} A {\bf 44}, 8024
(1991). 

\bibitem{perelomov}A.M. Perelomov, {\em Commun. Math. Phys.} {\bf 26}, 222 (1972).

\bibitem{bargmann}V. Bargmann, {\em Commun. Pure Appl. Math.} {\bf 14}, 187
(1961). 

\bibitem{deenen}J. Deenen and C. Quesne, {\em J. Math. Phys.} {\bf 25}, 2354 (1984);
D.J. Rowe, {\em ibid.} {\bf 25}, 2662 (1984).

\bibitem{sharma}J.K. Sharma, C.L. Mehta and E.C.G. Sudarshan, {\em J. Math. Phys.} {\bf
19}, 2089 (1978); J.K. Sharma, C.L. Mehta, N. Mukunda and E.C.G. Sudarshan,
{\em ibid.} {\bf 22}, 78 (1981).

\end{thebibliography}
\end{document}